# Saliency-Guided Deep Learning Network for Automatic Tumor Bed Volume Delineation in Post-operative Breast Irradiation

Mahdieh Kazemimoghadam[1], Weicheng Chi[1], Asal Rahimi[1], Nathan Kim[1], Prasanna Alluri[1], Chika Nwachukwu[1], Weiguo Lu[1] and Xuejun Gu[1]

1. Department of Radiation Oncology, University of Texas Southwestern Medical Center, Dallas, TX.
E-mail: xuejun.gu@utsouthwestern.edu
weiguo.lu@utsouthwestern.edu

**Abstract**

Efficient, reliable and reproducible target volume delineation is a key step in the effective planning of breast radiotherapy. However, post-operative breast target delineation is challenging as the contrast between the tumor bed volume (TBV) and normal breast tissue is relatively low in CT images. In this study, we propose to mimic the marker-guidance procedure in manual target delineation. We developed a saliency-based deep learning segmentation (SDL-Seg) algorithm for accurate TBV segmentation in post-operative breast irradiation. The SDL-Seg algorithm incorporates saliency information in the form of markers' location cues into a U-Net model. The design forces the model to encode the location-related features, which underscores regions with high saliency levels and suppresses low saliency regions. The saliency maps were generated by identifying markers on CT images. Markers' location were then converted to probability maps using a distance-transformation coupled with a Gaussian filter. Subsequently, the CT images and the corresponding saliency maps formed a multi-channel input for the SDL-Seg network. Our in-house dataset was comprised of 145 prone CT images from 29 post-operative breast cancer patients, who received 5-fraction partial breast irradiation (PBI) regimen on GammaPod. The 29 patients were randomly split into training (19), validation (5) and test (5) sets. The performance of the proposed method was compared against basic U-Net. Our model achieved mean (standard deviation) of 76.4(±2.7) %, 6.76(±1.83) mm, and 1.9(±0.66) mm for DSC, HD95, and ASD respectively on the test set with computation time of below 11 seconds per one CT volume. SDL-Seg showed superior performance relative to basic U-Net for all the evaluation metrics while preserving low computation cost. The findings demonstrate that SDL-Seg is a promising approach for improving the efficiency and accuracy of the on-line treatment planning procedure of PBI, such as GammaPod based PBI.

## 1. Introduction

Post-operative breast radiotherapy is a critical component in the multi-modal treatment of breast cancer. Following breast conserving surgery, radiotherapy is often employed to eradicate residual cancer using whole breast or partial breast irradiation (PBI) approaches. Whole breast radiation is often followed by a tumor bed boost to further reduce local recurrence risk (Yang *et al* 2013). In current clinical practice, tumor bed volume (TBV) delineation is performed manually on CT images prior to radiotherapy (RT) treatment planning. Manual contouring is tedious and one of the most time-consuming part of treatment-planning procedure. It is also user dependent, and thus prone to significant inter- and intra-subject variability (Geets *et al* 2005, Jensen *et al* 2014). Therefore, an automatic and accurate breast TBV segmentation algorithm is desirable to accelerate the target volume delineation process, improve the consistency of the results, and support an efficient clinical workflow. An accurate and efficient delineation is especially needed in online PBI treatment, such as Gamma-Pod-based PBI, where CT simulation, planning and treatment are completed in a single session (Yu *et al* 2013).

In recent years, deep learning (DL) models have received considerable attention in segmenting the target in radiotherapy (Meyer *et al* 2018, Liu *et al* 2017, Yang *et al* 2020). Men et al. developed a novel deep convolutional neural network (CNN) for auto-segmentation of tumor in planning the radiotherapy of rectal cancer (Men *et al* 2017). Guo et al. developed a 3D Dense-Net for gross-tumor volume (GTV) segmentation in head and neck multimodality images (Guo *et al* 2019). In (Wang *et al* 2018, Cardenas *et al* 2018), a deep CNN was leveraged to segment tumor boundary in nasopharyngeal carcinoma on MRI and CT images, and the results were shown to have comparable performance to manual segmentation. A novel deep dilated residual network for whole breast target delineation was also introduced in (Men *et al* 2018). Moreover, DL models have been widely applied to breast tumor detection and segmentation on diagnostic images (e.g., ultrasound and digital mammography images) (Huang et al 2017, Vakanski et al 2020, Khalili et al 2019, Dhungel et al 2015). For instance, Yap et al. proposed the application of DL for breast tumor segmentation in ultrasound images using three different methods including a patch-based CNN, a fully convolutional, and a transfer learning approach (Yap et al 2018). Zhang et al. developed a semi-supervised algorithm which integrates clinically-approved breast lesion characteristics into a DL model for automatic segmentation of ultrasound images (Zhang et al 2020). Full resolution convolutional network (FrCN) was also developed for breast tumor segmentation in digital X-ray mammograms (Al-Antari et al 2020). However, TBV delineation for breast radiotherapy treatment planning remains a challenge due to low contrast of TBV relative to the surrounding normal breast tissue as well as large variability in target size, shape, and location across patients. The inferior quality of planning CT scans particularly in soft tissue, low contrast to noise (CNR), and artifacts caused by high-density markers result in even further complexity of post-operative breast TBV delineation (Petersen et al 2007) (Yang et al 2013).

Among variety of DL models, U-Net (Ronneberger et al 2015) and its various modifications have received considerable attention for segmentation tasks in radiotherapy. In (Fu et al 2020) a U-Net based network was introduced for pelvic multi-organ segmentation and demonstrated promising results for accommodating prostate adaptive radiotherapy treatment planning. Chi et al developed a weakly-supervised approach using deformable image registration to train a 3D U-Net model for head and neck organs-at-risk segmentation (Chi et al 2020). Jin et al introduced a workflow for esophageal GTV and clinical target volume (CTV) segmentation by incorporating the strengths of U-Net architecture in propagating high-level information to lower-level (Jin et al 2021). A novel framework with recursive segmentation strategy was also developed by Chen et al for auto-segmentation of organs at risk using ensemble of two 3D U-Nets (Chen et al 2019a). Integrating prior domain knowledge into image segmentation models has been reported to improve model performance (Nosrati and Hamarneh 2016). According to a study by Chen et al, integrating prior anatomy information including tumor shape and positioning information into a convolutional neural network assist the DL model to accurately segment cervical tumors (Chen et al 2019b). Intensity, texture, and local features were also leveraged to successfully localize tumor regions and perform automatic segmentation of breast tumor (Xi et al 2017). In (Vakanski et al 2020) discriminative features were recognized by introducing anatomic prior knowledge and topological information into the DL segmentation model which led to improved model performance.

Post-operative patients with breast cancer often have markers (i.e., surgical clips, implanted fiducial markers) around the TBV. Such markers are visible in CT images (Jelvehgaran et al 2017), and can provide visual clues for TBV localization, hence assisting physicians in defining TBV with higher accuracy and consistency (Coles et al 2011), (van Mourik et al 2010) compared to the cases with no markers. Inspired by the visual location cues provided by the markers, we propose to develop a visual saliency guided DL model (SDL-Seg) for post-operative TBV segmentation. The proposed model aims to learn feature representations that prioritize spatial regions with high saliency levels and suppressing regions of low saliency to achieve high segmentation accuracy. To our knowledge, this is the first model to integrate domain knowledge for post-operative TBV segmentation. In this report, we discuss the implementation details of the proposed method. We then evaluate the model performance and report qualitative and quantitative segmentation results and compare them with those resulted from basic U-Net.

## 2. Methods and materials

### 2.1. SDL-Seg model

Figure 1 illustrates the proposed SDL-Seg model. The model is composed of three main steps: 1) Saliency maps generation: where the probability maps are generated by identifying markers as location cues to provide domain knowledge in the form of visual saliency in CT images; 2) Saliency-guided U-Net deep learning: in this step the saliency information is incorporated to guide the DL-based model to focus on image regions with high saliency values for an accurate segmentation of breast TBV. The CT images and the corresponding saliency maps form a multi-channel input for the first convolution layer of the SDL-Seg; 3) Majority voting: the predictions of multiple SDL-Seg networks are combined in this step to generate the final segmentation.

The probabilistic map of markers' spatial locations in CT images are utilized to encode the location of the target. During training, each image and its probability map are jointly fed into a U-Net model. Supervised by manually delineated contours, U-Net learns to look for the location cues with high intensity in the saliency maps and extract discriminative features. Network weights are continuously updated until the model tends to be stable. At the testing stage, CT images and their associated saliency maps go through the trained network. A thresholding is performed on all the predicted masks using a softmax function where the probability of the predicted voxels are set to either zero (background) or one (target).

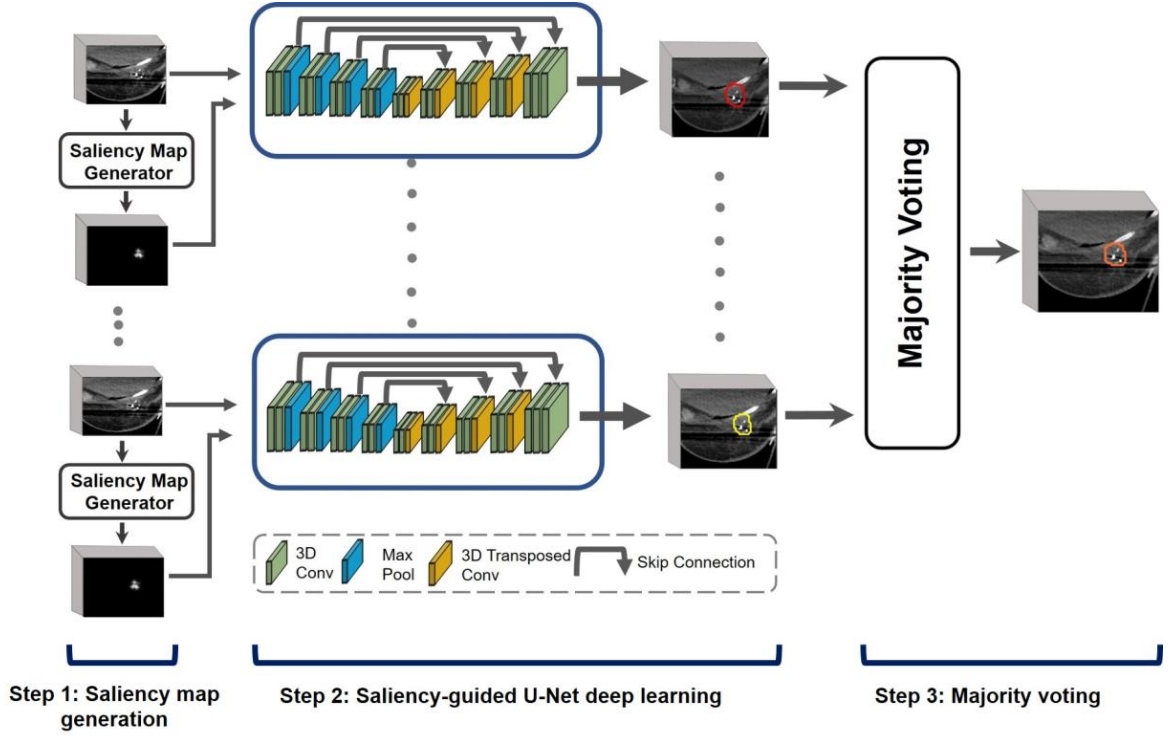

**Figure 1**. Illustration of SDL-Seg structure. The model consists of three main steps: 1. Saliency map generation where saliency maps with target location cues are produced, 2. Saliency-guided U-Net encoding the regions with high saliency for more accurate segmentation of breast target. 3. Majority voting fuses predictions of multiple U-Nets to generate final segmentation

### 2.1.1. Saliency maps

Our proposed saliency maps are inspired by the markers' visual cues for TBV localization. The level of saliency of image regions are described as how likely each region belongs to the target. In CT images, the intensity of markers is higher than that of their surrounding breast soft tissue. Masks with location cues were generated by empirically defining an intensity range that encompasses the markers. This was conducted by thresholding the CT images, where intensity values of 1 (markers) were assigned to the voxels with Hounsfield (HU) above 300, and 0s assigned to the background. The intensity of the ribs is similar to that of the markers which leads to unwanted regions in the resulting binary image. In order to eliminate ribs' intensity interference, whole breast masks were used to filter out the regions outside the breast tissue. Morphological operation steps including closing and opening were then employed to further refine the resulting binary image. This way non-marker structures were removed and the voxels belonging to the marker structures became fully connected. The number and location of the identified marker by the algorithm were manually validated to ensure accurate marker localization. Subsequently, Euclidian distance transformation coupled with a Gaussian filter was applied to convert markers' locations to probability maps. DGF (.) in (1) is a distance transformation coupled with a Gaussian filter. $\sigma$ controls the area surrounding each marker. In this study, $\sigma = 1$ voxel was selected as it provided the best results. For an input image, the output of such process is a visual saliency map where every voxel has a value ranging 0 to 1. Saliency values indicate the probability of a voxel belonging to the location cues.

$$DGF = \frac{e^{-Dst(p)^2}}{\sigma^2} \quad (1)$$

$Dst(p) = \min(ED(p, q_i))$ represents the Euclidian distance map (Danielsson 1980) between the voxel p and the nearest nonzero voxel q of the binary mask. DGFT in (2) is the normalized combination of DGFs across all location cues. n denotes the total number of location cues in the image.

$$DGFT = \frac{\sum_{j=1}^{n} DGF(j)}{argmax(DGF(1),...,DGF(j))} \qquad (2)$$

An example of a breast CT image with the markers enclosing the tumor bed and the corresponding saliency map is illustrated in Figure 2.

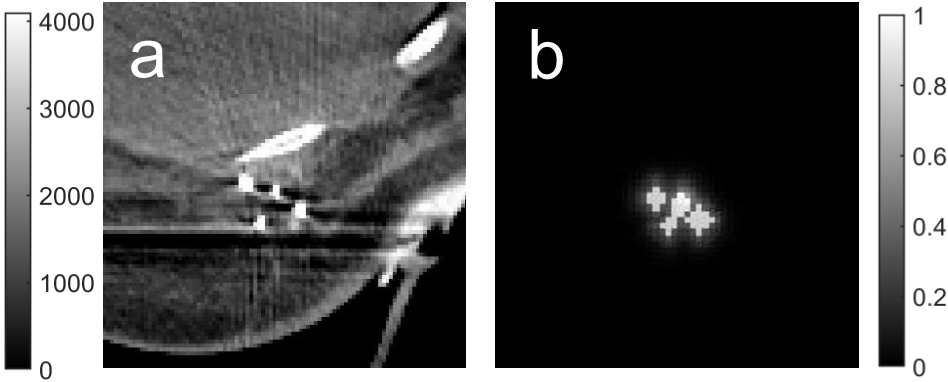

**Figure 2**. a) An example of a breast CT image with markers enclosing tumor bed. b) The corresponding saliency map.

### 2.1.2. Saliency-guided U-Net

We employed U-Net (Ronneberger *et al* 2015) as the backbone of our proposed method which consists of a contracting path and an expanding path. The contracting or down-sampling path is used to extract contextual information while expanding or up-sampling path enables localization and recovers spatial information. The contracting path is comprised of four convolutional blocks each followed by a 3D max pooling with the kernel size of 2. The first block had 32 channels. The number of channels is doubled after every max-pooling layer. Expanding path is also composed of four convolutional blocks to up-sample feature maps. The number of channels is halved in every step of the expanding path. Skip connections are applied to concatenate feature maps of the contracting path with its corresponding feature maps of expansive path. All the convolutional blocks include two consecutive convolutional layers with kernels of size 3×3×3, each followed by a batch normalization layer, and a Sigmoid activation function. A 3D spatial dropout with the rate of 0.2 was also applied after the second convolution layer of each block.

To incorporate saliency information into the network, CT images and the corresponding saliency maps formed a multi-channel input for the segmentation model. Such design forces the model to encode the location-related features, thus guiding the network to focus on regions with high saliency levels. The proposed model parameters are optimized by minimizing the loss function which compares the predicted contours with manually delineated ground truth labels. We used a combination of binary cross-entropy and Dice loss as the loss function. This hybrid loss leverages the advantages of both loss functions (Zhu *et al* 2019, Wong *et al* 2018). The Dice loss handles the imbalanced distribution of foreground and background voxels and cross entropy loss offers smooth gradient. The hybrid loss function is defined as:

$$L(G,P) = -\frac{1}{N}\sum_{k=1}^{K}\sum_{n=1}^{N}\left(g_{n,k} \log p_{n,k} + \frac{2\, g_{n,k}\, p_{n,k}}{g_{n,k}^2 + p_{n,k}^2}\right); \quad g_{n,k} \in G \quad p_{n,k} \in P \qquad (3)$$

Where G denotes the ground truth label, and P is the predicted probabilities for class k and voxel n in the image. An identical loss function was used to update the weights of the baseline U-Net during training.

### 2.2. Experimental data

To evaluate the proposed method, we collected 29 post-operative breast cancer patients from our institution to create an in-house dataset. All patients had markers in the form of either surgical clips or fiducials and received 5-fraction PBI regimen on GammaPod. Head-First-Prone (HFP) CT scans were acquired prior to each fraction on either Airo Mobile Intraoperative CT (Brainlab AG, Munich, Germany) or Philips Brilliance Big Bore CT (Philips Healthcare, Amsterdam, Netherlands), both with 120kVp, and scanning lengths cover the whole chest region. CT images had slice thickness of 1 mm. Pixel resolution varied between 1.17 mm and 1.37 mm. Contours of TBV were manually delineated by an attending physician on each treatment day on planning CT using Eclipse® treatment planning system (TPS) (Varian Medical Systems, Palo Alto, CA), and were defined as the ground truth labels for the segmentation task. Five fractions of each patient were contoured by 2-4 different physicians.

Out of 29 patients, 5 patients were randomly selected as the test set, and the remaining 24 patients were randomly split into training (19) and validation (4). Training, validation and test sets were comprised of a total of 95, 20, and 25 images respectively. Four-fold cross validation was applied to train the segmentation model. In order to improve the stability and accuracy of the final prediction, majority voting was then applied to combine the four predictions into the final segmentation. Validation data was used for tuning network's hyperparameters and selecting the best model. Test data was untouched during training and used to evaluate the final tuned model during the final evaluation of the segmentation performance.

### 2.3. Implementation details

For conducting experiments and implementing our segmentation model, we used Pytorch (1.8) DL library and python (version 3.7) environment on Windows 10 64x, Intel Xeon processor CPU with 64 GB RAM, and NVIDIA GeForce 2080 Ti GPU with 12 GB memory. To maintain consistency across all training data, CT images and the corresponding ground-truth labels were resampled to voxel size of 2mm×2mm×2mm. Images and labels were then cropped to 96×96×96. The voxel intensity of CT images was normalized to 0–1 according to HU window [−200 200]. The network was trained with the Adam optimizer and a learning rate of $10^{-4}$, batch size of 1, and maximum epochs of 200. The models performed best on validation sets were selected to evaluate the test dataset.

### 2.4. Network evaluation

We evaluated the performance of SDL-Seg model on our clinical breast CT dataset and compared it with the basic U-Net as the baseline algorithm. In section 3.1, the predicted contours by the SDL-Seg and basic U-Net were qualitatively evaluated with respect to the manually delineated contours. In section 3.2, we used Dice similarity coefficient (DSC), 95 percentile Hausdorff distance (HD95), and average symmetric surface distance (ASD) (Jafari-Khouzani *et al* 2011) to quantitatively asses the segmentation performance of our model and the baseline algorithm. DSC was computed to measure the degree of overlap between the SDL-Seg and basic U-Net with the ground truth masks. ASD and HD95 were computed to compare the distances between the models' segmentation output and the ground truth.

### 3. Results

### 3.1. Qualitative evaluation

Examples of input CT images with the ground truth masks (blue), the segmentation output by the proposed SDL-Seg (red) as well as the basic U-Net (yellow) are illustrated in Figure 3. The columns represent predicted masks for each fold as well as majority voting where the final segmentations were produced by fusing the predictions over folds. For the CT images displayed in Figure 3, the segmentation outputs by the basic U-Net were inferior compared to the predicted masks by SDL-Seg. In rows a, b, c and d, U-Net mistakenly produced negative predictions of TBV presence for image regions that belonged to the target. This is especially noticeable in row a, fold 2, where U-Net totally failed to locate the TBV region. The proposed SDL-Seg benefits from the information in the salient maps which guides the network to learn more detailed information of target location and shape priors and assist the network to encode the regions with high saliency. This led to great improvement of SDL-Seg relative to predictions by basic U-Net in rows a-d. In addition, rows e and f provide examples where U-Net erroneously segmented image regions of the neighboring tissue as TBV. This mainly occurred where basic U-Net could not handle the similarity of normal soft tissue intensity to the target. SDL-Seg significantly solved this issue by applying higher weights to the discriminative location cues provided by the saliency maps and focusing on the region relevant for the segmentation of the target. Hence effectively suppressed regions outside the TBV and further refined the predicted masks. Moreover, majority voting appeared to further assist in preserving the details of TBV boundary as well as producing more accurate segmentations. Majority voting was shown to reduce the negative effects of mis-predicted regions of the folds on the final segmentation. For instance, in case a, folds 2 and 4, a large area of normal tissue was mistakenly delineated as TBV, however, in the majority voting mask those normal tissue regions were excluded and the predicted TBV matched the ground truth very well. Fold 4 in case b also included negative predictions of TBV for image regions that belonged to the target, while in majority voting mask such erroneous predictions were significantly eliminated.

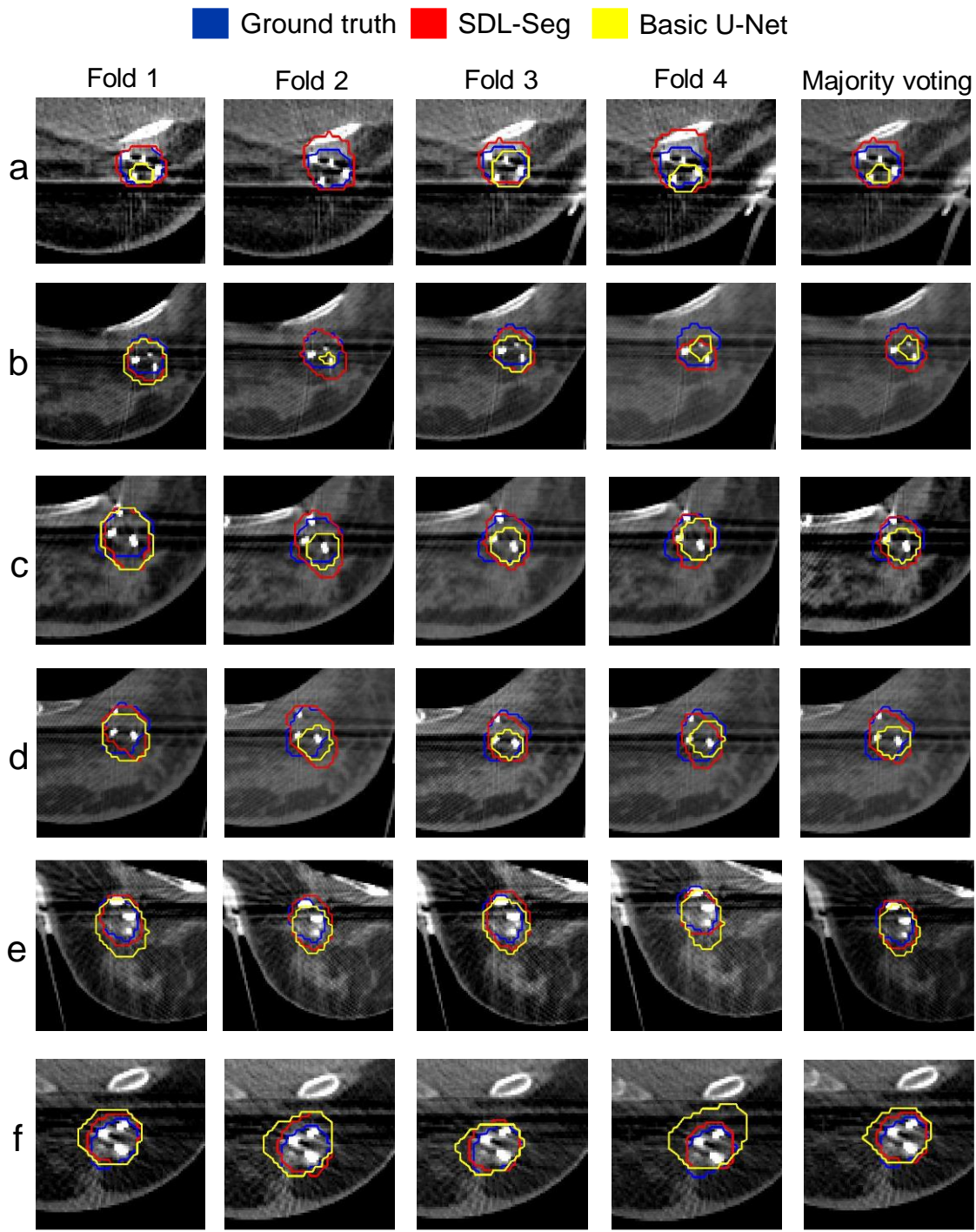

**Figure 3**. Qualitative comparison of the predicted breast target contour by the proposed SDL-Seg (red) and the segmentation by the basic U-Net (yellow) with respect to TBV contour (blue). The columns represent predicted masks for each fold as well as majority voting. Rows a-f are examples of the test cases.

## 3.2. Quantitative comparison

Table 1 summarizes the DSC, HD95 and ASD comparisons between the proposed SDL-Seg and basic U-Net for 5 test cases as well as the average and standard deviation across cases. SDL-Seg outperformed basic U-Net for all evaluation metrics. Mean DSC of 76.4% was achieved using SDL-Seg which outperformed basic U-Net by 13.8%. Average HD95 reduced by 1.63 mm using the proposed model relative to basic U-Net. Average ASD of 1.9 mm was observed for SDL-Seg outperforming basic U-Net by 0.9mm. Higher DSC was achieved using SDL-Seg relative to basic U-Net in all test cases by 3.5% to 21.7%, with ASD decreasing by 0.46 mm to 1.8 mm in SDL-Seg compared to basic U-Net. A reduction in HD95 ranging 0.32 mm to 5.32 mm was also observed using SDL-Seg for 4 test cases. Only in one test case, HD95 appeared to be lower by 1.4 mm using U-Net versus SDL-Seg.

**Table 1**. Quantitative comparison between the segmentation results of the proposed SDL-Seg and basic U-Net for the five test cases and the average across cases. Values represent mean ± standard deviation.

| Metrics | Methods | Case 1 | Case 2 | Case 3 | Case 4 | Case 5 | Mean |
|---|---|---|---|---|---|---|---|
| DSC (%) | U-Net | 70.8±2.1% | 57.8%±21.1% | 63.2%±4.3%, | 51.9%±9.5% | 69.2±3.7% | 62.6±7.8 |
| | SDL-Seg | **74.3%±4.8%** | **79.35%±3.7%** | **75.6%±1.7%** | **73.6%±4.4%** | **79.3%±1.5%** | **76.4±2.7** |
| HD95 (mm) | U-Net | 5.12±1.19 | 13.02±12.01 | 7.77±1.5 | 7.4±1.84 | 8.86±2.12 | 8.39±2.9 |
| | SDL-Seg | **4.8±1.07** | **7.7±10.37** | **9.196±1.02** | **5.12±1.48** | **6.93±2.43** | **6.76±1.83** |
| ASD (mm) | U-Net | 2.34±0.26 | 2.73±1.43 | 3.13±0.38 | 3.94±0.87 | 1.92±0.29 | 2.8±0.77 |
| | SDL-Seg | **1.88±0.37** | **1.27±0.1** | **2.89±0.21** | **2.12±0.56** | **1.34±0.19** | **1.9±0.66** |

## 3.3. Computational efficiency

Table 2 shows the comparison of computation time between basic U-Net and SDL-Seg on test dataset. The proposed SDL-Seg and basic U-Net accomplished the segmentation task for one typical CT image within relatively similar computation time (10.19 seconds and 10.24 seconds, respectively) implying that SDL-Seg does not cause extra computation burden to the baseline U-Net.

**Table 2.** Comparisons of average segmentation time of one typical CT image from the test set. Data handling refers to loading original NIFTI data from the hard drive and writing the prediction contour data back onto hard drive.

| Methods | Data Handling time (s) | Computation time (s) | Total time (s) |
|---|---|---|---|
| U-Net | 3.58 | 6.66 | 10.24 |
| SDL-Seg | 3.59 | 6.60 | 10.19 |

## 3.4. Coefficient of variation (CV)

In order to assess the level of variability of the segmentation results by SDL-Seg and the basic U-Net, CV was calculated across DSC of five fractions within each test case (Table 3). CV for cases 2-5 decreased significantly (by 3.42% - 31.83%) using SDL-Seg compared to basic U-Net. Only in case 1, lower CV was observed using U-Net versus SDL-Seg (2.98% vs 6.55%). Overall, CV was reduced by 9.72% using SDL-Seg indicating lower variability and higher consistency across predicted contours by SDL-Seg relative to basic U-Net.

**Table 3.** Coefficient of variation (CV) % across dice scores of five fractions for each test case using SDL-Seg and the basic U-Net.

| Coefficient of variation (CV) % | | | | | | |
|---|---|---|---|---|---|---|
| Methods | Case 1 | Case 2 | Case 3 | Case 4 | Case 5 | Mean |
| U-Net | 2.98 | 36.48 | 6.90 | 18.31 | 5.43 | **14.02** |
| SDL-Seg | 6.55 | 4.65 | 2.32 | 5.98 | 2.01 | **4.3** |

## 4. Discussion and conclusions

### 4.1. SDL-Seg versus basic U-Net

To accommodate an efficient scan-plan-treat clinical workflow in partial breast irradiation (PBI), fast, accurate and automated target delineation is desirable. Due to high variability of TBV in shape and size and low contrast of TBV boundary in CT images, identification of TBV is particularly difficult (González Sanchis *et al* 2013). The proposed saliency-guided method attempts to guide segmentations of U-Net by incorporating domain knowledge. The saliency maps indicate regions with markers, which attract visual attention and guide the segmentation network for better performance. Experiments on our in-house dataset demonstrated the promising segmentation capability of the proposed method over basic U-Net. SDL-Seg outperformed basic U-Net for all the performance metrics including DSC, HD95, and ASD by 13%, 1.78 mm and 0.9 mm respectively. Furthermore, CV of 4.3% and 14.02% were reported across DSC of the predicted contours by SDL-Seg and basic U-Net respectively, indicating higher stability and efficiency of SDL-Seg in TBV segmentation relative to the basic U-Net.

Scarcity of annotated data is a general issue in medical image segmentation. For the specific task of TBV delineation, labeled data is rare in public and building a dedicated large dataset demands lots of resources from radiation oncologists. By localizing target using marker's saliency information, the proposed SDL-Seg was shown to provide robust performance despite the relatively small training dataset.

### 4.2. Impact of saliency map's quality

The segmentation results of the proposed SDL-Seg appeared to highly rely on the quality of the saliency maps. The visualization of some fiducial markers can be more difficult than others due to differences in marker's intensity. The visibility and contrast of markers in CT images and how well they can be identified are highly dependent on the marker size and the material they are made of (Habermehl *et al* 2013). These factors along with the number of markers appeared to affect the quality of the generated saliency maps in our study. For instance, in our dataset, gold markers were observed to have an excellent contrast. Using CT images containing gold fiducials for generating saliency information led to good-quality maps, which contained sufficient location cues, thus significantly improving outcomes compared to basic U-Net. In contrast, maps with sparse and small location cues such as when fewer small-size markers are identified, cannot provide adequate information for the segmentation model. Such saliency information results in either degraded performance or, at best, no improvement of the segmentation. Figure 4 demonstrates an example where the impact of saliency map's quality was studied on the model performance. We started with a saliency map with one location cue and increased the number of location cues in each step (a(1) to a(5)). The generated saliency maps were incorporated into the DL model and DSC was calculated for the predicted contour with respect to the ground truth. Up to five location cues were incorporated in total. Saliency maps with low number of markers mistakenly segmented areas of TBV as background resulting in large erroneous prediction (Figure 4). Increased number of location cues enriches useful saliency information, which is the underlying reason for more accurate predictions. According to the results, the predicted contour expands towards the newly added marker in each step. Therefore, in order to guide the proposed DL model for better performance, one potential approach could be to incorporate physician's input where saliency information may not be sufficient. For instance, physicians can expand location cues by placing virtual landmarks on CT images, so that saliency information is further refined. Future research should also consider developing robust and automatic algorithms for identifying a range of commercially available markers of varied sizes and material to generate high quality saliency maps. During marker identification process, we observed that having prior information on the marker type (e.g., marker material, size) and their quantity could be helpful in marker localization. Such prior information could assist to reduce human error during manual verification of the markers in the CT images especially for small size markers.

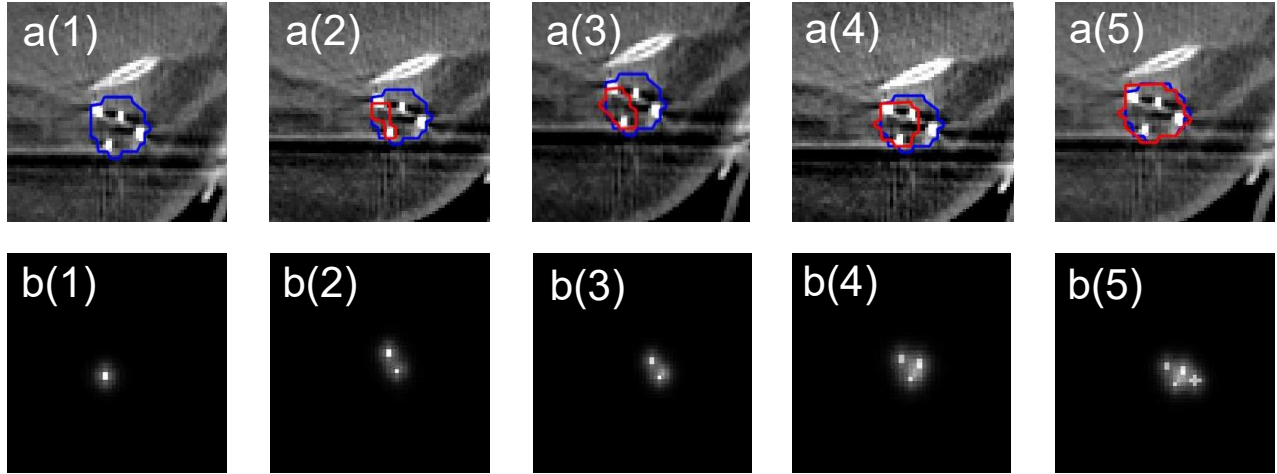

**Figure 4.** The effects of saliency map's quality on the segmentation outcomes. a(1)-a(5): Examples of CT images with TBV contour (blue) and the predicted contour by the proposed saliency-guided structure (red). b(1)-b(5): Saliency maps with varying levels of location cue information.

### 4.3. Variation of manually delineated contours

Figure 5 summarizes the coefficient of variation (CV) calculated across TBV volumes of five fractions for each case. The results indicate a wide range (CV=5-38%) of inter-observer variation in manual contouring of planning CT images, with an average CV of 14.35% across cases. The segmentation performance of our proposed model provided DSC of 76.4% and is comparable to human experts', considering the significant inter-observer variability between experts' manual segmentations of TBV. On the other hand, cases with high variation of ground truth data could adversely affect model training and result in degraded performance of the segmentation algorithm (Tajbakhsh *et al* 2020). In order to mitigate such effects, approaches such as identifying weakly delineated samples during training and lowering their impact by excluding them from backpropagation (Min *et al* 2019) should be considered in future studies.

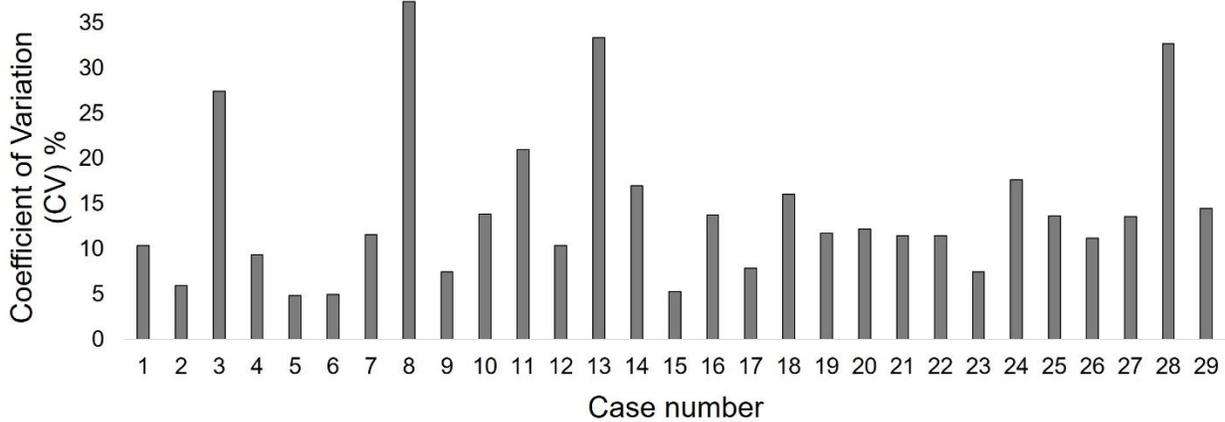

**Figure 5**. Coefficient of variation (CV) across five manually delineated TBV contours for each case.

An intra-observer variability study was also performed where one fraction from each test case was re-contoured by the same physician. The breast target volume during planning and after re-contouring, as well as intra-observer dice score were calculated for each test case (Table 4). According to Table 4, the average intra-observer dice score across 5 test cases is 82.7%± 5.4. Comparing this results with the outcomes by SDL-Seg (Table 1) indicates that the predicted masks by the proposed method are comparable to intra-observer variability in TBV segmentation.

|  | Case1 | Case 2 | Case 3 | Case 4 | Case 5 |
|---|---|---|---|---|---|
| Target volume during planning (cc) | 7.74 | 5.76 | 8.85 | 8.54 | 8.98 |
| Target volume after re-contouring (cc) | 8.77 | 6.44 | 10.14 | 8.45 | 8.48 |
| Dice score (%) | 75.8 | 82.6 | 85.8 | 79.5 | 89.8 |

In summary, we developed a saliency-guided DL model, SDL-Seg, integrating visual saliency in the form of markers' location into a U-Net model for TBV delineation. Results demonstrate that SDL-Seg outperforms basic U-Net while preserving computation cost. This is a promising approach for improving the efficiency and accuracy of the on-line treatment planning procedure for breast radiotherapy.